\pdfoutput=0
\documentclass[twocolumn]{aastex6}
\usepackage{graphicx}
\usepackage{epsfig}
\usepackage{amsmath, amssymb , amsthm, fouriernc}
\usepackage{threeparttable}

\def\lapp{\ifmmode\stackrel{<}{_{\sim}}\else$\stackrel{<}{_{\sim}}$\fi}
\def\gapp{\ifmmode\stackrel{>}{_{\sim}}\else$\stackrel{>}{_{\sim}}$\fi}

\shorttitle{{\it NuSTAR} Observations of Magnetar 1E~1048.1$-$5937}
\shortauthors{Yang et al.}

\begin{document}

\title{{\it NuSTAR} Observations of Magnetar 1E~1048.1$-$5937}

\author{C. Yang\altaffilmark{1,2}, R. F. Archibald\altaffilmark{2}, J. K. Vogel\altaffilmark{3}, H. An\altaffilmark{4}, V. M. Kaspi\altaffilmark{2}, S. Guillot\altaffilmark{5}, A. M. Beloborodov\altaffilmark{6}, M. Pivovaroff\altaffilmark{3}}

\altaffiltext{1}{Beijing Institute of Technology, 5 South Zhongguancun Street,Haidian District, Beijing 100081,China}
\altaffiltext{2}{Department of Physics \& McGill Space Institute, McGill University, 3600 University St., Montreal, QC, H3A 2T8, Canada}
\altaffiltext{3}{Lawrence Livermore National Laboratory, PLS/Physics, Livermore, CA 94550, USA}
\altaffiltext{4}{KIPAC, Stanford University, Stanford, CA 94305-4060, USA}
\altaffiltext{5}{Instituto de Astrofisica, Facultad de Fisica, Pontificia Universidad Catolica de Chile, Av. Vicuna Mackenna 4860,
782-0436 Macul, Santiago, Chile}
\altaffiltext{6}{Physics Department and Columbia Astrophysics Laboratory, Columbia University, 538 West 120th Street, New York, NY, 10027, USA}

\begin{abstract}
We report on simultaneous {\it NuSTAR} and {\it XMM-Newton} observations of the magnetar
1E~1048.1$-$5937, along with {\it Rossi X-ray Timing
Explorer (RXTE)} data for the same source. 
The {\it NuSTAR} data provide a clear detection of
this magnetar's persistent emission up to 20 keV.
We detect a previously unreported small secondary peak in the average
pulse profile in the 7--10~keV band, which grows to an amplitude comparable
to that of the main peak in the 10--20 keV band.  We show using {\it RXTE}
data that this secondary peak is likely transient.
We find that the pulsed fraction increases with
energy from a value of $\sim$0.55 at $\sim$2~keV to a value of $\sim$0.75 near 8~keV but
shows evidence for decreasing at higher energies.  
After filtering out multiple bright X-ray bursts
during the observation \citep{akb+14}, we find that the phase-averaged
spectrum from combined {\it NuSTAR} and {\it XMM} data is well described by
an absorbed double blackbody plus power-law model, with no evidence
for the spectral turn-up near $\sim$10~keV as has been seen 
in some other magnetars.   Our data allow us to rule out a spectral turn-up
similar to those seen in magnetars 4U 0142+61 and 1E 2259+586 of
$\Delta\Gamma \gapp 2$, where $\Delta\Gamma$ is the difference between the
soft-band and hard-band photon indexes.  The lack of 
spectral turn-up is consistent with what has been observed from an 
active subset of magnetars given previously reported trends suggesting
the degree of spectral turn-up is correlated with spin-down rate and/or spin-inferred magnetic field.
\end{abstract}


\section{Introduction}
Magnetars have hard X-ray spectra that represent an interesting puzzle.
Previously considered to be `soft' X-ray sources
\citep[e.g.][]{ms95,vtv95}, with energy spectra falling rapidly below $\sim$10~keV owing
to a steep ($\Gamma \sim -3$ to $-4$) power-law photon index,
the discovery of hard X-ray emission from several magnetars
\citep{rsv+04,klm04,khdc06}, and the realization that their spectra can {\it rise} in
energy above 10~keV, clearly demonstrated that in some cases, most
of the energy in persistent magnetar emission arises in the hard X-ray band.

With now 8 magnetars exhibiting observable high energy emission
\citep[see][and references therein]{ok14}, some possible trends have been noted.
The pulsed fraction of the hard X-ray emission has been suggested to rise
with energy to reach $\sim$100\% by $\sim$100~keV \citep{khdc06} although more recent
studies using the {\it Nuclear Spectroscopic Telescope Array} ({\it NuSTAR}) have suggested otherwise for
one source \citep{ahk+13}.  The spectra of many magnetars appear to continue
rising above $\sim$100~keV \citep{khdc06,dkh08} although evidence for a spectral turnover near 300~keV
was detected using {\it INTEGRAL} in the magnetar 4U 0142+61 \citep{hkw08}.
\citet{kb10} noticed a possible anti-correlation between the
degree of the soft--hard spectral turnover and either the frequency derivative
or the magnetic field strength \citep[see also][]{enk+10}.

The physical origin of magnetar hard X-ray emission is not yet well understood.
Several models have been proposed in the literature
\citep{hh05,bh07,tb05,bt07} although each has been argued to be problematic 
\citep[see, e.g.][for a discussion]{kb10}.
Most recently, \citet{bel13} proposed a model that explains the hard X-ray
emission as being due to escaping radiation from photons scattered by
relativistic particles in the outer regions of large active magnetic
loops called `j-bundles'.
This model has been successfully applied to {\it NuSTAR} detections
of magnetars 1E~1841$-$045 \citep{aah+15}, 1E~2259+586 \citep{vhk+14}
and 4U~0142+61 \citep{thy+15}, as well as other observations \citep{hbd14},
yielding interesting constraints on source geometries.

1E~1048.1$-$5937 is the only persistently bright magnetar (here `bright' is defined as
quiescent 2--10~keV unabsorbed flux $ > 5 \times 10^{-12}$~erg~cm$^{-2}$~s$^{-1}$) not yet
to have shown clear persistent hard X-ray emission.  Though monitored
for many years with the {\it Rossi X-ray Timing Explorer (RXTE)} Proportional Counter
Array (PCA) \citep[e.g.][]{dkg09,dk14}, and in spite of the PCA's excellent
sensitivity to hard X-rays albeit with no focusing capability, 1E~1048.1$-$5937
was not detected in detailed analysis by \citet{khdc06}, although
\citet{lwr08} did report a marginal detection of unpulsed flux (4.5$\sigma$)
in {\it INTEGRAL} data in the 22--100~keV band.

{\it NuSTAR} observed 1E~1048.1$-$5937 in 2013 in order to search for persistent
hard X-ray emission.  In this 320-ks observation, eight bright 
X-ray bursts were detected fortuitously; these have been reported on by \citet{akb+14}.
Although bursts from this source have been previously detected
\citep[e.g.][]{gkw02,gk04,dk14}, they are rare (five bursts were detected in
2.9 Ms for an average burst rate of one per week, assuming bursts are produced 
randomly, consistent with the behavior of this source). Hence the detection of eight 
bursts in 320~ks was somewhat unexpected, although in many magnetars bursts
are seen to be clustered \citep[e.g.][]{kgw+03}.  

It spite of the bursting behavior, the pulsar's persistent
emission during the {\it NuSTAR} observation is known from independent 
{\it Swift} X-ray Telescope monitoring observations \citep{akn+15} to have been
nominal.

Here we report on these same 2013 {\it NuSTAR} observations of 1E~1048.1$-$5937,
this time focusing on the persistent hard X-ray emission.  We detect, for the first
time, pulsations above $\sim$10~keV, only the 9th such magnetar detection yet.  We show that these
pulses are detected as high as 20~keV but no higher, precluding the same dramatic spectral turn-up
observed in some other magnetars.\footnote{Note that well after our submission of this paper,
    we were made aware of the work of \citet{wg15} who also analyzed these same data
and found similar results.}

\section{Instruments and Observations}
\label{sec:obs}

In order to study the pulsar's hard X-ray emission, we conducted simultaneous observations using
{\it NuSTAR} and ESA's {\it X-ray Multi-Mirror Mission} 
({\it XMM-Newton}). 

{\it NuSTAR} is a NASA Small Explorer (SMEX) satellite mission operating in the $3-79$~keV range \citep{hcc+13}.  This is the first hard X-ray satellite mission (above $\sim$10 keV) with focusing capability. The instrument features two focal plane modules, dubbed FPMA and FPMB. Each module consists of a reflective, multilayer-coated, focusing telescope \citep{hab+10,cab+11} with a CdZnTe detector sitting in the focal plane \citep{hchm10}. The achievable FWHM energy resolution varies from $400$~eV at $10$~keV to $900$~eV at $68$~keV, while the angular resolution of the instrument is characterized by a half-power diameter (HPD) of 58$\arcsec$ and a FWHM of 18$\arcsec$. 

The {\it XMM-Newton} observatory~\citep{jla+01} focuses on the soft X-ray regime ($0.1-10$~keV)
using its three on-board X-ray telescopes. Each of these Wolter I grazing-incidence optics consists of $58$ coaxially and co-focally nested mirrors. A European Photon Imaging Camera (EPIC) sits at the prime focus of each of the three telescopes. Two of these Charge-Coupled Devices (CCDs) make use of the MOS technology (MOS-1/MOS-2), while the third detector is a pn-CCD \citep{taa+01,sbd+01}.
The achievable energy resolution lies between $70-80$~eV and the EPIC point-spread function is $5-6\arcsec$ FWHM ($15\arcsec$ HPD).

{\it NuSTAR} observed 1E 1048.1$-$5937 in July of 2013.  A total of four separate observations took place between July 17 and July 27 ({\it NuSTAR} observation IDs 30001024002, 30001024003, 30001024005 and 30001024007). As mentioned above,  in these observations, which together comprised a total of 320-ks of exposure, eight bright X-ray bursts were detected \citep{akb+14}. For the analysis presented in this paper we used the same data set, but excluded times of burst activity (bursts and their tails) to avoid contaminating timing and spectral results with burst emission.

\begin{figure}[]
\begin{centering}
\includegraphics[width=0.5\textwidth]{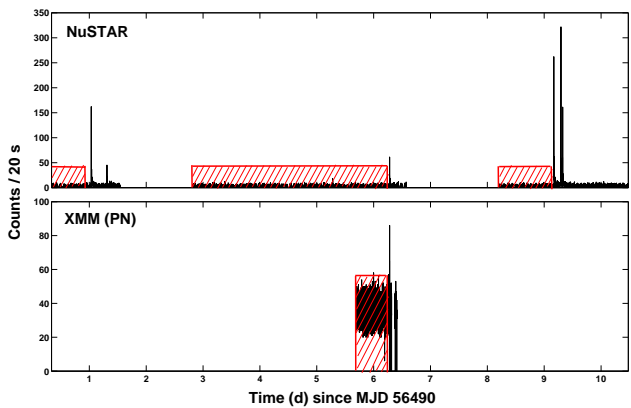}
\caption{Light curve of {\it NuSTAR} observation (top) with combined FPMA and FPMB data and {\it XMM-Newton} observation (bottom) displaying the EPIC-pn data. Red regions indicate good time intervals after excluding the bursts and their tails.}
\label{fig:lc}
\end{centering}
\end{figure}

The selected portions of the light curve are shown in red in the
top panel of Figure~\ref{fig:lc}.  Note that only 3 of the 4 {\it
NuSTAR} data sets could be used in this analysis as the fourth was
too contaminated by bursts and their tails \citep[see][]{akb+14}.
The corresponding good time intervals (GTI) used in our analysis lie
between MJDs $56490.33 - 56490.93$, MJDs $56492.81 - 56496.19$ and MJDs
$56498.19 - 56499.14$. The total exposure time of all GTIs is $210$~ks
for the presented analysis.  Table~\ref{ta:observations} provides details on
all data sets used in this analysis.

\begin{table*}
\scriptsize
\begin{center}
\caption{Summary of \emph{NuSTAR}, \emph{XMM-Newton} and \emph{Swift} observations of 1E 1048.1$-$5937 used in this study.}
\label{ta:observations}
\vspace{4mm}
\begin{tabular}{ccccccc}
\tableline\tableline
Observatory & Instrument & Mode& ObsID & Date [MJD]\tablenotemark{a} & Date [yymmdd]\tablenotemark{a} & Exposure [ks]  \\
\tableline
\emph{NuSTAR} & $\cdots$        & $\cdots$ & 30001024002   & 56490.33 & 130717  & 24.7/24.5\tablenotemark{b} \\
\emph{NuSTAR} & $\cdots$        & $\cdots$ & 30001024003   & 56490.97 & 130717  & 24.0/0.0\tablenotemark{b}   \\
\emph{NuSTAR} & $\cdots$        & $\cdots$ & 30001024005   & 56492.83 & 130719  & 156.7/140.7\tablenotemark{b} \\
\emph{NuSTAR} & $\cdots$        & $\cdots$ & 30001024007   & 56498.21 & 130725  & 111.3/47.4 		         \\
\emph{XMM}    & EPIC-pn         & Full Frame & 0723330101  & 56495.69 & 130722  & 48			         \\
\emph{XMM}    & EPIC-MOS1/2     & Small Window & 0723330101& 56495.67 & 130722  & 64  			          \\
\emph{RXTE}      & PCA                & Good Xenon   &  see \S\ref{sec:rxte}  & 52630--54200 & 021222--070410  & 939.5  \\
\emph{RXTE}      & PCA                & Good Xenon   & see \S\ref{sec:rxte}   & 54500--56000 & 080204--120314  & 1101.0  \\
\tableline
\end{tabular}
\tablenotetext{a}{At the start of data acquisition.}
\tablenotetext{b}{Before/after burst data removed, except for the {\it RXTE}/PCA data, which contained no bursts.}
\end{center}
\end{table*}

{\it XMM-Newton} observed 1E 1048.1$-$5937 from July 22-23, 2013
(Observation ID 0723330101), simultaneous with part of the {\it NuSTAR}
observations, as shown in Figure~\ref{fig:lc}. Our {\it XMM} observation
made use of the EPIC pn-CCD as well as the MOS-type detectors \citep{taa+01,sbd+01}. The
total exposure time for the EPIC pn-CCD observations 
was $48$~ks, while the MOS1/2 data set included $64$~ks.
All cameras were acquiring in imaging mode with the pn-CCD using Full
Frame Mode and MOS1/2 detectors running in Small Window Mode.

We processed the {\it NuSTAR} data using the standard {\it
NuSTAR} Data Analysis Software  {\tt NuSTAR\-DAS} version
1.4.1, including the standard data processing routines
{\tt nupipeline} and {\tt nuproducts}, with the HEASoft software
package\footnote{http://heasarc.nasa.gov/lheasoft/} (version 6.16) and
CALDB\footnote{http://heasarc.gsfc.nasa.gov/docs/heasarc/caldb/caldb\_intro.html}
(version 20140414).  To include the GTIs in our data screening we made
use of the {\tt usrgtifile} parameter for the {\tt nupipeline} and {\tt
nuproducts} commands. In the subsequent analysis, source events were
selected from within a circular region of 60$\arcsec$ radius at the {\it
Chandra} position of 1E 1048.1$-$5937 \citep{ok14}, while background
events were extracted within a circular region of 100$\arcsec$ radius
far away from the source.  The barycentric correction to the selected
events was done using the multi-mission tool command {\tt barycorr}
together with the appropriate orbital and clock correction files.

In addition to the phase-averaged spectral analysis described below in
\S\ref{sec:phaseavg}, we conducted a phase-resolved study (\S\ref{sec:phaseres}).  For this
purpose, we defined MJD 56200.00002243577 as the time of phase
zero and calculated the phase of each event in both the source and
background region using the command {\tt dmtcalc} in the CIAO software
package\footnote{http://cxc.harvard.edu/ciao/} (version 4.6).  We then
employed the XSELECT tool of the HEASoft package to extract a spectrum
in each of the chosen phase ranges.  We regrouped the source spectrum to
have a minimum of 30 counts per energy bin and conducted all following
spectral analyses using XSPEC (version 12.8.2), which is also part of
HEASoft. 
Because there were no significant source counts above
20~keV, we limited our analysis to the energy range of $3-20$~keV in
the {\it NuSTAR} data.
For all our spectral analyses, we assumed interstellar absorption
cross sections from \citet{bm92} \citep[with updates from][]{ysd98} 
and abundances from \citet{ag89}.

To process the {\it XMM-Newton} data we made use of the {\tt SAS}
data analysis software package\footnote{http://xmm.esac.esa.int/sas/}
(version 13.5.0), and employed the most recent calibration files
(downloaded July 16, 2014)
and standard threads\footnote{http://xmm.esac.esa.int/sas/current/documentation/threads/} to process the
data. In order to avoid pile-up issues we
selected an annular region around the source with an inner and outer
radius of $3\arcsec$ and $30\arcsec$, respectively. The background was
selected in a source-free circular region of $60\arcsec$ in radius. The
{\it XMM-Newton} light-curve (Fig.~\ref{fig:lc}, bottom) shows that there
was a burst during the {\it XMM} observation, which was also observed
in the {\it NuSTAR} data as reported by \citet{akb+14}. For the analysis
described in this paper, only the {\it XMM-Newton} data before the burst
were used.

To obtain the phase-resolved spectrum for {\it XMM-Newton}, we used the
{\tt barycen} command to apply the barycenter correction. The phase of
each event was then obtained using {\tt phasecalc}. Both commands are
part of the standard SAS package. Following these steps we extracted
a spectrum for each phase bin with the {\tt evselect} command in the
energy range $0.3-10$~keV and regrouped the data to obtain a minimum
of $30$~counts per spectral bin.

Because of the bursting seen in the {\it NuSTAR} data, it is conceivable
that tail emission following bursts that may have occured prior to the start of our 
three data intervals (see Fig. 1) could contaminate the results that follow.  
Given the short duration
of tails seen following 1E~1048.1$-$5937 bursts \citep{gkw06,akb+14}, 
even if there had been a bursts prior to all three data intervals, 
at most 4\% of our data would contain tail emission.
Nevertheless, we have verified by examining archival data from {\it Swift}/BAT and {\it Fermi}/GBM 
that those instruments detected no bursts from the direction of
1E~1048.1$-$5937 during the relevant epochs.  
Additionally, we have verified that there is no evidence for a decreasing
count rate at the start of any of the three observing intervals, as would be expected
if tails were present there.  Given the absence of a decreasing count rate and
the properties of the bursts described by \citet{akb+14}, notably how the
spectrum evolves with time, our
simulations of the worst-case contamination and find our fitted spectral
parameters for the persistent emission could not have been significant affected.

\section{Analysis and Results}
\label{sec:results}

\subsection{Timing Analysis}
\label{sec:timing}

\begin{figure}[t!]
\begin{centering}
\includegraphics[width=0.53\textwidth]{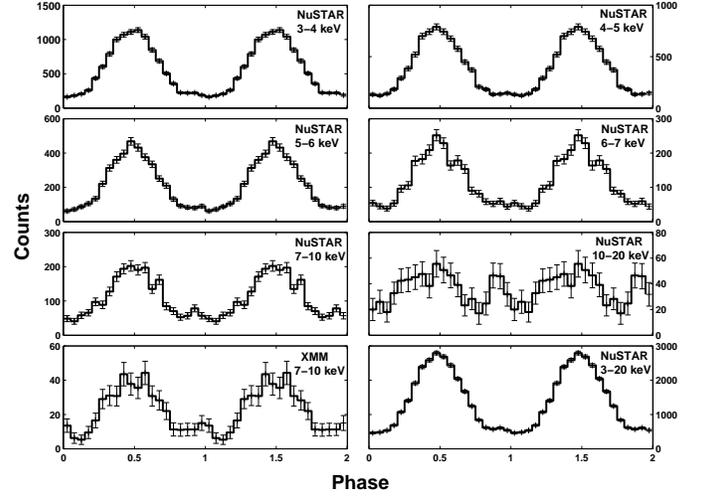}
\caption{{\it NuSTAR} pulse profiles in various energy bands after folding with the best-fit period and after
excluding the bursts and their respective tails.  The full 3--20 keV profile is shown the bottom right plot,
and the {\it XMM} 7--10 keV profile is shown the bottom left plot, for comparison with the {\it NuSTAR}
7--10 keV plot.  Note the secondary peak that appears in this 7--10 keV band, and which grows in relative
amplitude in the 10--20 keV band.}
\label{fig:profile}
\end{centering}
\end{figure}

For the timing analysis, we searched for pulsations in the $3-20$ keV barycenter-corrected, event list using the H-test \citep{dsr89}
and determined the best period for 1E 1048.1$-$5937 in the {\it NuSTAR} data to be $6.4616815(3)$~s. This period is consistent with that found in contemporaenous monitoring observations with the {\it Swift} X-ray Telescope \citep{akn+15}.
We folded the barycenter-corrected events with this best-fit period to obtain the pulse profiles in different energy ranges for {\it NuSTAR} and {\it XMM-Newton} data separately.  The results are shown in Figure~\ref{fig:profile}.

Below $10$~keV, the pulse profiles are largely sinusoidal in shape with only one single distinctive peak, in agreement with previous observations \citep[e.g.][]{tmt+05}. Interestingly, there is a small secondary peak in the 7--10 keV {\it NuSTAR} pulse profile, consistent with a small excess apparent in the pulse profile from the {\it XMM} data in the same energy range, as shown in the bottom left panel of Figure~\ref{fig:profile}. For the first time, pulsations above 10~keV are observed in this source.  The secondary peak from the 7--10 keV band grows in amplitude such that the pulse profile of 1E~1048.1$-$5937 exhibits two peaks in the 10--20 keV energy band. 
This secondary peak has a harder spectrum than the first and is roughly out of phase by 180$^{\circ}$, consistent with being from an opposite pole.  However, as described below
(\S\ref{sec:rxte}) this secondary peak is not detected in {\it RXTE} data from past epochs and so appears
to be transient.
Above about $20$~keV, no pulsations were detected, presumably due to a paucity of counts.\\

We also calculated the pulsed fraction (PF) as a function of energy for our {\it NuSTAR} and {\it XMM} observations using the folded pulse profiles and following the prescription described in \citet{aah+15}. Specifically, we calculated
the area pulsed fraction PF$_{\rm area}$ (the difference between pulsed flux and constant flux integrated over a full phase cycle)
and the RMS pulsed fraction PF$_{\rm RMS}$ (a measure of the deviation of the pulsed flux from its mean) 
using 6 harmonics.  Note that PF$_{\rm area}$ and PF$_{\rm RMS}$ are expected to differ as they measure different quantities; see
\citet{aah+15} for details.

\begin{figure}[]
\begin{centering}
\includegraphics[width=0.5\textwidth]{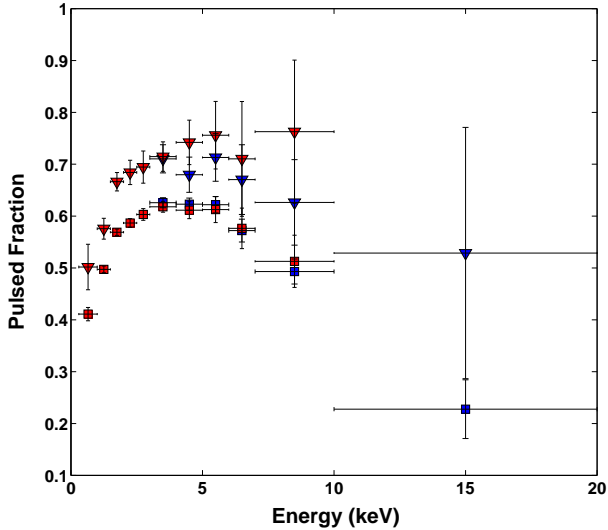}
\caption{Area (triangles) and RMS (squares) pulsed fractions as a function of energy for the {\it NuSTAR} and {\it XMM} (EPIC-pn) pulse profiles. {\it NuSTAR} data are shown in blue, {\it XMM} data in red.}
\label{fig:pf}
\end{centering}
\end{figure}

Figure~\ref{fig:pf} shows the results of our pulsed flux analysis for
the {\it NuSTAR} and {\it XMM} (EPIC-pn) pulse profiles, specfically
PF$_{\rm area}$ and PF$_{\rm RMS}$ as a function of energy.  In the
Figure, {\it NuSTAR} data are shown in blue, while red refers to {\it XMM}
(EPIC-pn) data. The area pulsed fractions and the RMS pulsed fractions
are displayed as triangles and squares, respectively. The energy bands for
the {\it NuSTAR} data correspond to those shown in Fig.~\ref{fig:profile},
while for the {\it XMM} data we used the 0.3--10 keV band.

The observed trend for both PF$_{\rm area}$ and PF$_{\rm RMS}$ is a slight
increase with energy up to about $5$~keV and then a decrease
for increasing energies.  The latter is not statistically
significant for PF$_{\rm area}$ considering the large uncertainties especially in the largest
two energy bins, however it is for PF$_{\rm RMS}$.  For energies around $1$~keV, PF$_{\rm area}$
is $50\% \pm 4\%$, which rises to about $74\% \pm 4\%$ for energies around
$5$~keV and ``drops" to $53\% \pm 24\%$ for $10-20$~keV. The results for
PF$_{\rm RMS}$ are $41\% \pm 1\%$, $61\% \pm 2\%$, and $23\% \pm 6\%$
for energies around $1$, $5$, and $10-20$~keV, respectively.

\subsection{Phase-Averaged Spectral Analysis}
\label{sec:phaseavg}

To study the spectrum of 1E~1048.1$-$5937, we combined
data from {\it NuSTAR} and {\it XMM}.  The {\it XMM} data were useful for establishing the low-energy spectrum, particularly
the thermal component, since {\it NuSTAR} is not sensitive below 3~keV.  
We used only pn {\it XMM} data for the spectral analysis because of the fewer MOS counts
and to minimize cross-calibration systematic errors.  For pn we chose only data in the energy interval 0.3--10 keV,
and for {\it NuSTAR} (FPMA+FPMB) only 3--20~keV data were used,
because above 20 keV there were insufficient source counts for a meaningful result.  We rebinned these data to have a minimum of 30 counts per spectral bin.  

\begin{figure}[]
\begin{centering}
\includegraphics[width=0.5\textwidth]{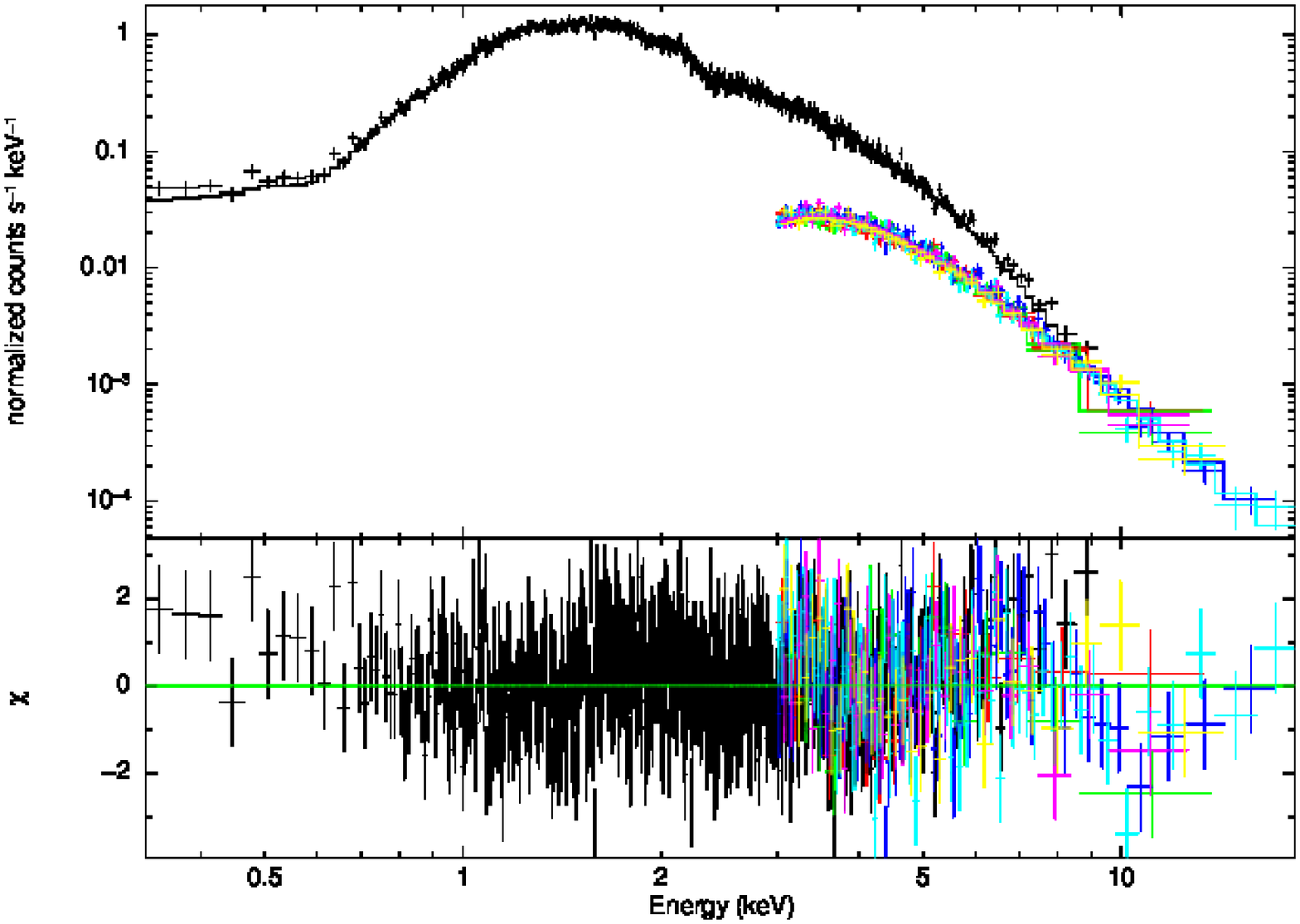}
\includegraphics[width=0.5\textwidth]{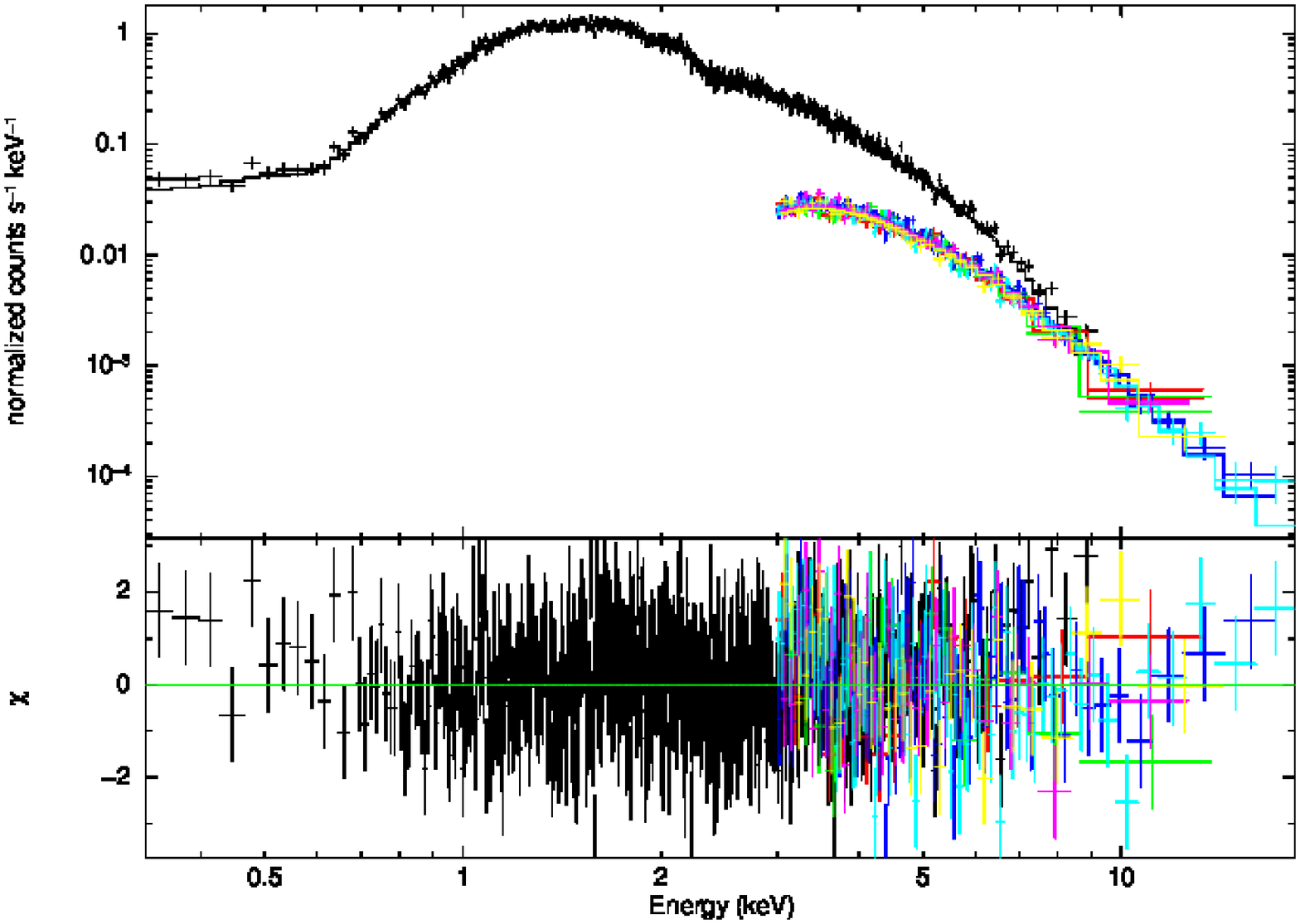}
\caption{Joint spectral fitting results using {\it NuSTAR} (multi-colored) and {\it XMM}/pn (black) data, shown such that each spectral bin has a minimum of 80 counts. We used all the good-time-interval {\it NuSTAR} data in the range 3--20 keV, and {\it XMM}/pn data in the range 0.3--10 keV. Top:  Fit and residuals for an absorbed blackbody plus power-law model.  Bottom:  Fit and residuals for a model consisting of an absorbed blackbody plus a broken power law. See Table 2 for fit details.}
\label{fig:spectra}
\end{centering}
\end{figure}

We began with a blackbody plus power-law model, often employed for magnetars, and used
the $\tt{tbabs*const*(cflux*bbody+cflux*powerlaw)}$ model in XSPEC.   We held the pn relative normalization at unity, 
and let the normalization values of the {\it NuSTAR} data groups vary. 
This absorbed blackbody plus power-law model yielded best-fit parameters of absorption $N_{H} = 1.22\pm0.01\times10^{22}$~cm$^{-2}$, blackbody temperature 
0.667$\pm$0.005 keV, and photon index 3.64$\pm$0.02.  Our best-fit $N_H$ is significantly
higher than that reported by \citet{tgd+08} based on joint fits of {\it Chandra, Swift} and {\it XMM} data
in the energy range 0.5--7~keV.  The different energy ranges and different instruments used may account
for the discrepancy; we note that our best-fit $N_H$ is statistically consistent with that measured
by \citet{tmt+05} (their Observation C) which was based on {\it XMM} data only, and included data
up to 10 keV.   
We tried different abundance and cross-section models (see \S\ref{sec:obs}) but
these did not make significant differences to the fit, and resulted in
$N_H$ values that differed from the above but at most a few percent.
For our best-fit model, the 3--10 keV keV flux
in the blackbody component is $(2.03 \pm 0.03) \times 10^{-12}$~erg~s$^{-1}$~cm$^{-2}$, and in the power-law component
$(1.78 \pm 0.03) \times 10^{-12}$~erg~s$^{-1}$~cm$^{-2}$.
The reduced $\chi^2$ to the fit described above was 1.07 with 1352 degree of freedom which has probability $\sim$4\% of
being due to chance.  This low probability is suggestive -- though not conclusive -- 
that the model is not optimal.
Figure~\ref{fig:spectra} (top) shows the spectrum and residuals after subtraction of this simple model; indeed the residuals have some structure that suggests this model is not optimal.

\begin{table*}
\vspace{0.0 mm}
\begin{center}
\caption{Phase-averaged spectral parameters for joint {\it NuSTAR} and {\it XMM} spectral fits.$^a$}
\label{ta:spectra}
\scriptsize{
\begin{tabular}{ccccccccccccc} \hline\hline
$Model^b$ & $N_{H}$ & $kT1$ & $kT2$ & $\Gamma_{s}$ & $E_{break}$ & $\Gamma_{h}$ & $\chi^2_{red}$/dof  \\
  & ($10^{22}$ cm$^{-2}$) & (keV) & (keV) &  & (keV) &   &  \\ \hline
BB+PL & 1.22(1) & 0.667(5) & $\cdots$ & 3.64(2) & $\cdots$ & $\cdots$ & 1.07/1352  \\
BB+BknPL & 1.14(2) & 0.624(8) & $\cdots$ & 3.36(5) & 6.3(2) & 4.4(1) & 1.03/1350  \\
2BB+PL & 1.17(2) & 0.53(3) & 0.85(5) & 3.67(5) & $\cdots$ & $\cdots$ & 1.03/1350  \\
2BB+BknPL & 1.22(5) & 0.51(2) & 0.87(5) & 4.1(3) & 5.3(8) & 3.0(5) & 1.03/1348  \\
\hline
\end{tabular}}
\end{center}
\footnotesize{
$a$ Uncertainties are at the $1\sigma$ confidence level.  {\it NuSTAR} energy range is 3--20 keV with combined FPMA and FPMB data.  {\it XMM} energy range is 0.3--10 keV with pn data only.\\
$b$ BB: blackbody, 2BB: two blackbodies, PL: power law, BknPL: broken power law.}
\end{table*}

For this reason, we also tried fitting these data with multiple different
models.  Specifically, we employed an absorbed blackbody plus broken
power-law model to search for a spectral break, as is seen in 
some other magnetars.
This model yielded a reduced $\chi^2$ value of 1.03 (see Table 2).
Figure~\ref{fig:spectra} (bottom) shows residuals for the 
broken power-law model.
For completeness, we also tried a double blackbody plus power-law
model as has been used in other hard X-ray analyses of magnetars
\citep{hbd14}, as well as a
double blackbody plus broken power-law model, both with absorption.
These yielded similar goodness-of-fits as the blackbody plus broken power law.
Best-fit spectral parameters for all the models we tried
are provided in Table~\ref{ta:spectra}.  
We verified that the improvement of the fits using these models over the
simpler 2-component model was
statistically significant using simulations.  Using {\tt XSPEC}, we 
created fake datasets having simple blackbody plus power-law spectra 
with parameters equal to those found when fitting such a model (see above).
We then fit these fake data sets with the aforementioned multi-component models
(blackbody plus broken power law, double blackbody plus power law, double blackbody
plus broken power law)
and recorded the change in fit statistic.  For each trial model, in 1000 trials, in no case did
the fit statistic change by an amount even close to that measured for
the real data set.  We hence conclude with $>$99.9\% confidence that
the improvement in fit for the 3-component models is required by the data.

Importantly, for the blackbody plus
broken power-law model, the best-fit results in a spectral break at 6.3
keV, and a significant spectral {\it softening}, with power-law index
changing from 3.36 to 4.4.  This is quite different from what has been 
observed for other magnetars for which hard X-ray emission has
been detected \citep[see][for a summary]{ok14}.

\begin{figure}[]
\begin{centering}
\includegraphics[width=0.5\textwidth]{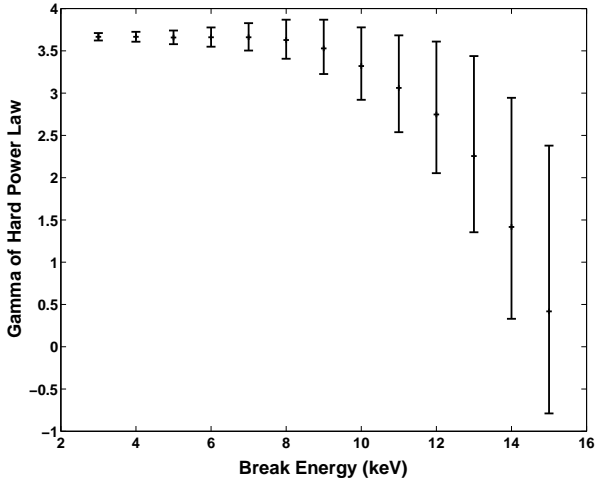}
\caption{Best-fit hard-band photon index $\Gamma_h$ as a function of assumed break energy.  In making this plot,
we held fixed the best-fit parameters from the absorbed double blackbody plus power-law model 
(see Table~\ref{ta:spectra}) in an absorbed double blackbody plus broken power-law model; we
held the break energy at integer values in keV ranging from 3 to 15~keV, each time fitting
for $\Gamma_h$ using the {\it XMM} and {\it NuSTAR} data jointly.}
\label{fig:break}
\end{centering}
\end{figure}

To investigate this possible softening further, and
because there exists strong covariance between the break energy and change in
photon index, we did the following analysis.  
Noting that a double blackbody plus power law provides a reasonable parameterization of
the data, we used the best-fit absorbed double
blackbody plus power-law model parameters (see Table~\ref{ta:spectra})
and kept them fixed while fitting an absorbed double blackbody plus
broken power-law model.  We then held the spectral break fixed at values
between 3 and 15 keV, at 1 keV intervals, each time fitting only for
the hard-band photon index, $\Gamma_h$.  The results are shown in Figure~\ref{fig:break}.
The $\Gamma_h$ fit in this way shows marginal evidence for spectral hardening,
rather than softening, however it is not statistically significant.  
We use this analysis instead to
estimate an upper limit on any spectral change.  Assuming a fiducial break
energy of 10~keV, we set a 3$\sigma$ limit on $\Gamma_h$ to be
$>$1.8, which limits any change in the power-law index from the soft band to be $\Gamma_s - \Gamma_h = 3.67 - 1.8 \equiv \Delta\Gamma < 1.9$.

We performed additional simulations to investigate this result and in particular ask
the question of whether we can rule out spectral turn-ups as have been
seen in other magnetars.  We took from the literature summarized by
\citet[][]{kb10} the spectra of magnetars 4U~0142+61, 1E~1841$-$045, 
1E~2259+586 and 1RXS J1708$-$4009, specifically their respective spectral
break energies and degree of spectral turn-up, as parameterized by
$\Delta\Gamma$.  On top of our best-fit absorbed double blackbody
plus power-law model, we  added a spectral break and generated fake spectra in XSPEC
using the {\tt fakeit} command.  The fake spectra were grouped to have at least 20 counts
per spectral bin. Spectral fits were then performed in the 3--79 keV band with 
double blackbody plus broken power-law and a double blackbody plus power-law models,
and the F-test probability was calculated given the results of every fit.  The F-test
probabilities (useful since we are comparing nested models) 
suggested that we can rule out with confidence
$\Gamma_s - \Gamma_h < 1.9$. This is consistent with our independent analysis above, which
gives us confidence we have found a robust limit.

In particular, we
rule out spectral turn-ups like those observed in 4U 0142+61 ($E_b = 12$~keV, $\Delta\Gamma=3.6$) 
and 1E~2259+586 ($E_b = 11.5$ keV, $\Delta\Gamma=2.9$) with
very high confidence (probability $9 \times 10^{-25}$ and $2 \times 10^{-15}$, respectively), 
that for 1RXS~J1708$-$4009 ($E_b = 16$ keV, $\Delta\Gamma=1.9$) at the 95\% confidence level,
but cannot rule out that for 1E~1841$-$045 ($E_b = 13.5$ keV, $\Delta\Gamma=0.7$) with any confidence (probability 0.14).
We thus conclude that we can rule out $\Delta\Gamma \gapp 2$ for 1E 1048.1$-$5937.

\subsubsection{Upper Limit on 20--79 keV Flux}
\label{sec:ul}

In addition, we calculated the $3\sigma$ upper limit on the total,
phase-averaged flux for 1E~1048.1$-$5937 in the $20-79$~keV range
using the conservative approach of \citet{pkc00}
to determine the count rate upper limit.
This yielded an upper limit 
of $7.7(3)\times 10^{-3}$~cts~s$^{-1}$.  We then used the NASA HEASARC tool
WebPIMMs\footnote{https://heasarc.gsfc.nasa.gov/cgi-bin/Tools/w3pimms/w3pimms.pl}
based on PIMMS version 4.8 to obtain an estimate on the flux upper limit
by using the values for hydrogen column density $N_H$ and the photon
index of the expected power law\footnote{We neglected the blackbody
contribution to the energy range from $20-79$~keV} (PL). This yielded
$3.3(1)\times 10^{-12}$~ergs~cm$^{-2}$~s$^{-1}$ using the McGill Online Magnetar
Catalogue\footnote{http://www.physics.mcgill.ca/~pulsar/magnetar/main.html}
($N_H=0.97(1)\times 10^{22}$~cm$^{-2}$, $\Gamma=3.14(11)$) and
$2.9(1)\times 10^{-12}$~ergs~cm$^{-2}$~s$^{-1}$ for the best-fit values determined
in our spectral analysis (see Table~\ref{ta:spectra}), i.e. $N_H=1.22(1)\times 10^{22}$~cm$^{-2}$,
$\Gamma=3.64(2)$.  To obtain a more robust upper limit on the $20-79$~keV
we then used the {\tt fakeit} command in XSPEC taking into account
the {\it NuSTAR} response, which yields the $3\sigma$ flux upper limit of
$4.15(13)\times10^{-12}$~ergs~cm$^{-2}$~s$^{-1}$ ($20-79$~keV) for our
best-fit values (Table~\ref{ta:spectra}).

\subsection{Phase-Resolved Spectral Analysis}
\label{sec:phaseres}

\begin{figure}[]
\begin{centering}
\includegraphics[width=0.5\textwidth]{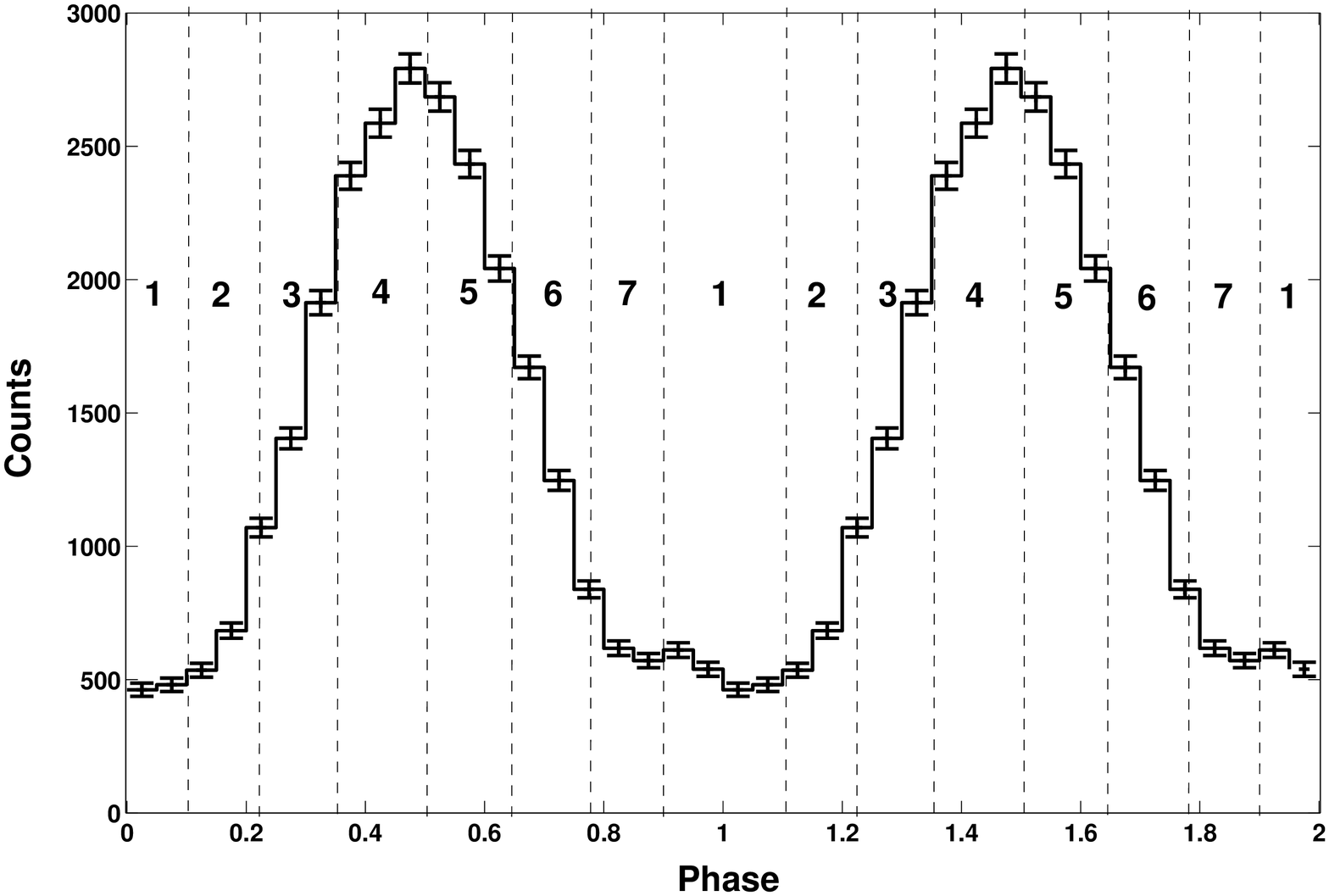}
\caption{The 3--20 keV pulse profile of our {\it NuSTAR} observation.   For phase-resolved spectroscopy, the pulse profile was divided into seven parts, shown using black vertical dashed lines. Phase 1 is 0.0--0.1 and 0.9--1.0, phase 2 is 0.1--0.225, phase 3 is 0.225--0.35, phase 4 is 0.35--0.5, phase 5 is 0.5--0.65, phase 6 is 0.65--0.775, and phase 7 is 0.775--0.9.}
\label{fig:phasedef}
\end{centering}
\end{figure}

\begin{figure}[]
\begin{centering}
\includegraphics[width=0.5\textwidth]{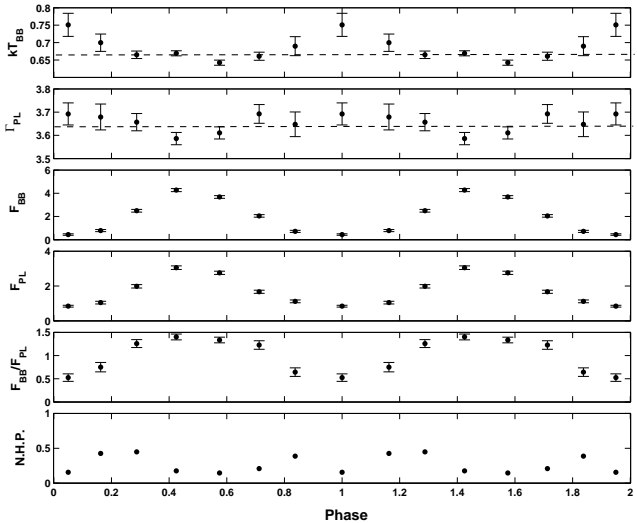}
\caption{Phase-resolved spectral parameters determined assuming an absorbed blackbody plus power-law model. From top to bottom: the blackbody temperature in units of keV, the power-law index, 3--10 keV unabsorbed flux of the blackbody component in units of $10^{-12}$~erg~s$^{-1}$~cm$^{-2}$, 3--10 keV unabsorbed flux of the power-law component in units of $10^{-12}$~erg~s$^{-1}$~cm$^{-2}$, the ratio of the blackbody flux to the power-law flux in the 3--10 keV band, and the probability of obtaining the best-fit reduced $\chi^2$ for the spectral fit.  The error bars indicate $1\sigma$ uncertainty ranges. Dashed lines represent best-fit phase-averaged values (see Table 2).} 
\label{fig:phaseBBPL}
\end{centering}
\end{figure}

We defined MJD 56200.00002243577 as phase zero and used the period provided by the H-test (see \S\ref{sec:timing}) to obtain folded light curves for the {\it NuSTAR} and {\it XMM} data in the 3--20 keV and 0.3--10 keV bands, respectively. 
We divided the folded light curves into seven parts, as shown in Figure~\ref{fig:phasedef}, echoing the
phases defined by \citet{tmt+05}.  We adopted an
absorbed blackbody plus power-law model to fit each phase bin.  
The best-fit absorption, $N_{H}$, $1.22\times10^{22}$~cm$^{-2}$, from the phase-averaged fitting (see \S\ref{sec:phaseavg}), was used and held fixed in 
the phase-resolved spectral analysis.  The simple absorbed blackbody plus power-law model fit all phase bins well.
The results are summarized in Figure~\ref{fig:phaseBBPL}, where we show the
variability of the model parameters with pulse phase.  
The blackbody temperature in this spectral parameterization is varying significantly; a $\chi^2$ test yields a
0.2\% probability of the observed variation being due to chance.
The blackbody temperature is minimal at pulse maximum, suggesting
spectral hardening away from the main pulse.   
On the other hand, in our data, the variation of the power-law index suggests the main pulse is harder than the off-pulse contribution, however a $\chi^2$ test indicates the variation is not significant,  with a 17\% chance of the variation being due to chance.
Our analysis finds that the
ratio of thermal to non-thermal flux is a maximum at the pulse peak.  We note that the secondary
peak we observe in the 10--20 keV band suggests spectral hardening away from the main peak, however at a level that
is small and hence difficult to quantify.

\subsection{{Long-term {\it RXTE} Observations of 1E~1048.1$-$5397}}
\label{sec:rxte}

In order to confirm the absence of a spectral turn-up in 1E~1048.1$-$5937 observed by {\it NuSTAR}, as well as to investigate the
secondary peak we detected in the 10--20 keV {\it NuSTAR} pulse profile,
we analyzed observations of 1E~1048.1$-$5937 from the Proportional Counting Array (PCA) aboard {\it RXTE}.
The PCA consists of five collimated xenon/methane multianode proportional Counter Units (PCUs) which are sensitive to photons in the 2--60$\;$keV range \citep{jsg+96,jmr+06}.
1E~1048.1$-$5937 was monitored with the PCA regularly for most of the lifetime of {\it RXTE}; see \cite{dk14} for a summary.
For all observations of 1E~1048.1$-$5937, the PCA was operated in ``Good Xenon'' mode, which provides 1-$\mu$s resolution for photon arrival times.
These observations were obtained from the HEASARC archive and reduced to the barycenter 
using the {\tt barycorr} tool in HEASoft version 6.16.
Observations were filtered to remove non-astrophysical events using {\tt xtefilt}.

As 1E~1048.1$-$5937 has experienced several long-term flux flares \citep[see][]{dk14, dkg09, gk04}, we selected only observations from MJDs 52630--54200 and from MJDs 54500--56000, two periods of time where the reported 2--20$\;$keV flux was relatively stable.
This resulted in 2.06 Ms of data with an average of 2.07 PCUs on.
Observations were folded using local ephemerides from \cite{dk14}, and then aligned by cross-correlating the profiles in the 2.0--5.5 keV band, where the signal-to-noise ratio is highest in these data.

\begin{figure}[]
\begin{centering}
\includegraphics[width=0.48\textwidth]{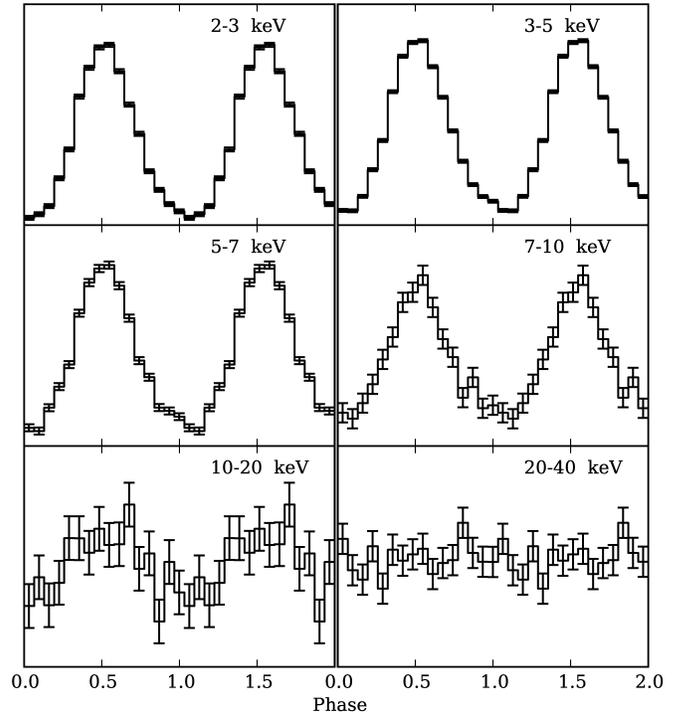}
\caption{{\it RXTE}/PCA pulse profiles for 1E~1048.1$-$5937 in various energy bands.  Note the absence
of a secondary peak in the 10--20 keV band, in contrast to that observed with {\it NuSTAR} (Fig.~\ref{fig:profile}).
Also note the absence of any pulsations in the 20--40 keV band.}
\label{fig:rxte}
\end{centering}
\end{figure}

In this data set, a pulsed signal is detectable at the lower-energy end of the PCA band and into the 10--20$\;$keV band.
PCA pulse profiles for various energy bands, including 10--20~keV, are shown in Figure~\ref{fig:rxte}.  Note the absence
of any evidence of a secondary peak in the 10--20 keV PCA profile, in contrast to that seen in the
{\it NuSTAR} data (see Fig.~\ref{fig:profile}).  This is also consistent with what was reported by \citet{khdc06}.
This demonstrates that this profile feature is likely
transient in nature.  Also note the apparent absence of any pulsations in the 20--40 keV PCA band.
There is a possible hint of a secondary feature in the 7--10 keV band, however it is not statistically significant.

\begin{table}[th]
\begin{center}
\caption{{\it RXTE}/PCA Pulsed Fluxes for 1E 1048.1$-$5937}
\label{ta:rxte}
\begin{tabular}{lcl} \hline\hline
Band & Pulsed Count Rate & Unabsorbed Pulsed Flux$^a$ \\
     & (Counts PCU$^{-1}$s$^{-1}$) & (erg cm$^{-2}$ s$^{-1}$) \\
\hline
\hline
2--3 keV & $0.0813\pm0.0005$ & ($4.84\pm0.03$)$\times10^{-12}$ \\
3--5 keV & $0.150\pm0.001$ & ($1.48\pm0.01$)$\times10^{-12}$ \\
5--7 keV & $0.060\pm0.001$ & ($5.38\pm0.09$)$\times10^{-13}$ \\
7--10 keV & $0.017\pm0.001$ & ($1.7\pm0.1$)$\times10^{-13}$ \\
10--20 keV & $0.005\pm0.001$ & ($0.2\pm0.25$)$\times10^{-13}$ \\
20--40 keV & $0.\pm0.002$ & $\textless 2$$\times10^{-13}$ \\ \hline
\end{tabular}
\end{center}
{\small
 $a$ Pulsed flux calculated using WebPIMMs assuming an absorbed power-law \\ 
plus blackbody with  $N_H=1.22\times10^{22}$cm$^{-2}$,  $\Gamma=3.64$, and $kT=0.667$.}
\end{table}

We present pulsed count rates for the different energy bands in Table~\ref{ta:rxte}.
To convert the {\it PCA} pulsed count rate into unabsorbed pulsed fluxes, we used WebPIMMs, 
assuming the phase-averaged spectral parameters we found in \S\ref{sec:phaseavg}.
These PCA count rates are consistent within uncertainties with the results from our {\it XMM}/{\it NuSTAR} analysis.

\section{Discussion \& Conclusions}
\label{sec:discussion}

Our joint {\it NuSTAR} and {\it XMM} data have provided a clear detection of
pulsed emission from 1E~1048.1$-$5937 up to 20 keV.
We have discovered a previously unreported small secondary peak in the average
pulse profile in the 7--10~keV band, which grows to an amplitude comparable
to that of the main peak in the 10--20 keV band.  {\it RXTE}/PCA data averaged
over several years prior to these new observations do not show any evidence
of the secondary peak, indicating it is transient. 
We also have shown that the pulsed fraction of 1E~1048.1$-$5937 increases with
energy from $\sim$2~keV to a value of $\sim$0.75 near 8~keV but
shows evidence for decreasing at higher energies.  
In our spectral analysis,
after filtering out multiple bright X-ray bursts
\citep{akb+14}, we have shown that the source's phase-averaged
spectrum is well described (though not uniquely) by
an absorbed double blackbody plus power-law model.  The data show no evidence
for the spectral turn-up near $\sim$10~keV seen 
in some other magnetars.  Indeed, for 1E~10481.1$-$5937, we can rule out a spectral turn-up
similar to those seen in magnetars 4U 0142+61 and 1E 2259+586 of
$\Delta\Gamma \gapp 2$.   We have also studied the phase-resolved spectrum and
have observed clear spectral changes with rotational phase.

It is important to compare the {\it NuSTAR}-constrained hard X-ray flux of 1E 1048.1$-$5937
with other estimates, in order to look for variability.
The marginal {\it INTEGRAL} detection of 1E~1048.1$-$5937 \citep{lwr08} in the 22--100 keV
band was reported without any spectral information, but with a count rate.  By comparing
the count rate with that reported for a different source ($\eta$ Carina) in the same
paper, and assuming the two sources have similar spectra in that energy band, we infer
a 22--100 keV flux of $\sim 5 \times 10^{-12}$~erg~s$^{-1}$~cm$^{-2}$.  
This is nominally just above our 3$\sigma$ upper limit in the 20--79~keV band
(see \S\ref{sec:ul}).  However given the uncertainties particularly in the {\it INTEGRAL}
value, the two fluxes cannot be considered inconsistent.  Nevertheless we may ask, given the soft X-ray variability
seen in the source \citep{dk14,akn+15}, whether the {\it INTEGRAL} or {\it NuSTAR/XMM} data
were taken at substantially different phases in the source's X-ray flux evolution.  If so,
the comparison of their hard X-ray fluxes, however crude, may not make sense.
The {\it INTEGRAL} data were
averaged over several years, mostly from May 2003 to June 2005.  During that time span, the
2--10 keV flux of 1E~1048.1$-$5937 was slowly declining following a bright flare in mid-2002 and
the source was undergoing rapid (yet unexplained) torque variations during the first half
of the interval \citep{gk04,dk14}.
On the other hand, the {\it NuSTAR/XMM} data were also taken during the decline of
the source flux following a flux flare at the start of 2012, but after the cessations
of the strong torque variability following that event \citep{akn+15}.  Hence the source
did exhibit somewhat different rotational behavior during the reported {\it INTEGRAL} observing
epoch and that of our observations, although this, and the uncertainty in the {\it INTEGRAL}
flux, are insufficient for concluding the hard X-ray flux varies. 
The consistency of the pulsed fluxes from {\it RXTE}/PCA (Table~\ref{ta:rxte}) with those
from our {\it NuSTAR/XMM} analysis strengthen our conclusion that we find no evidence for
flux variation in the hard band for this source, although this is not a strong conclusion.

The absence of an observed spectral turn-up is interesting and not unexpected in the
model of \citet{bel13}.  One possible explanation is unfavourable geometry.
If the 
object is viewed close to the rotation axis,
e.g. at 10$^{\circ}$--20$^{\circ}$, 
and the magnetic dipole axis is weakly inclined to the spin axis
\citep[as suggested by observations; see][]{hbd14}, then the line-of-sight to the emission remains 
close to the magnetic axis at all rotation phases. In this case
the predicted hard X-ray component is generally
weak \citep[see Fig. 7 in][]{bel13}.  In other words, the source may actually
produce copious hard X-rays, but we do not observe them due to an unfavourable
viewing angle.  
The geometry of an axisymmetric j-bundle viewed near
the magnetic axis may also be consistent with the reduction
of pulsed fraction at high energies (Fig.~\ref{fig:pf}).
On the other hand, for a small angle between the line-of-sight and
the rotation axis, a high pulsed fraction at low energies as is observed would not obviously
be expected.  It could be that the inclination is larger but we do not see the
j-bundle because it is not axisymmetric and instead confined to
a small range of magnetic azimuth, so that its field lines are never
tangent to our line of sight (and therefore do not emit hard X-rays
toward our direction).  Another possibility is that the object has a weak or
non-existing magnetic twist/j-bundle.  However, in this case it
is unclear why its emission extends to 20 keV.

The absence of an observed spectral turn-up is consistent with a trend noted by
\citet{kb10} between degree of spectral upturn and both spin-inferred magnetic field strength
and spin-down rate in magnetars.  The spin-inferred surface dipolar magnetic field strength of
1E~1048.1$-$5937 is $\sim 4 \times 10^{14}$~G, but because its spin-down rate can vary by
over a factor of 10 {\citep{akn+15}, $B$ inferred from this spin-down should be regarded with caution.
The trends noted by \citet{kb10} suggested that 1E~1048.1$-$5937 should have a hard-band
$\Gamma_h$ of 1--2, and that the difference between soft-band and hard-band photon indexes,
$\Gamma_s - \Gamma_h \equiv \Delta\Gamma$, should be 0--1.\footnote{Note, in the text of \citet{kb10} the authors
predicted $\Gamma_h \sim 0-1$, which is not consistent with their own trends.  We believe this
was an oversight.}  From the current data, we cannot constrain $\Gamma_h$ very strongly.  However we can rule out sharp
spectral upturns as observed in magnetars 4U~0142+61 and 1E~2259+586.  In this sense, our results for 1E~1048.1$-$5937
are consistent with the reported $\Gamma_s - \Gamma_h$ correlation.

The small secondary peak in the pulse profile seen in the 10--20 keV
band appears to be a new feature, as previous {\it RXTE} data \citep[see
Fig.~\ref{fig:rxte} and][]{kgc+01,khdc06} show no evidence for it.  This
may suggest that the source was not in a true quiescent state during the
{\it NuSTAR/XMM} observations, as other temporary features in the pulse
profile have been reported, notably near the epochs of bright flux flares
\citep[see][and references therein]{dk14}.  Indeed the bursting behavior
we detected during our observation is consistent with the source being
in some form of outburst.  On the other hand, at the epoch of our {\it
NuSTAR} and {\it XMM} data, as discussed above, the source had largely
recovered from its 2012 flux flare and was in a relatively rotationally
stable phase \citep{akn+15}.  In any case, the hard-band pulse profile,
in contrast to the source hard-band flux, is clearly variable.

In our phase-resolved analysis, we found that the blackbody temperature, as judged from the absorbed blackbody plus power-law model, was highest off-pulse.
This is {\it opposite} to the trend reported by \citet{tmt+05}, who found the
main pulse to have a higher blackbody temperature.   We note that the behavior in phase of the power-law index in
our analysis, namely a harder $\Gamma$ on-pulse, is consistent with that seen by \citet{tmt+05}, however not
to the same degree:  our power-law index varies by $\sim$0.1 over a period, while theirs did by $\sim$0.4.
However, an important caveat in this comparison is that in the \citet{tmt+05} analysis, no attempt was made
to vary the blackbody and power-law components simultaneously,
presumably due to lack of counts.  Hence, the comparison of the behaviors of either $kT$ or $\Gamma$
is not exact.  Regardless, both analyses find that the
ratio of thermal to non-thermal flux is a maximum at the pulse peak, with variation by a factor of $\sim$3 over
the period, although again, \citet{tmt+05} do not vary both model components, so the comparison is not ideal.

Our observations have served to continue to flesh out the hard X-ray emission properties of magnetars
as a population.  For the target in question, 1E~1048.1$-$5937, the source's faintness at hard X-ray
energies precludes detailed modelling using the framework of \citet{hbd14} as has been used in other
{\it NuSTAR} magnetar studies.  However, a comparably
long second joint {\it NuSTAR/XMM} observation could at least test whether the bursting fortuitously detected
in the first observation impacted the hard-band properties, particularly the appearance of the secondary
pulse peak.  An observation much longer than ours would be required to enable detailed physical modelling.
Long-term monitoring with the LAXPC instrument aboard the {\it Astrosat} mission has the potential
to reveal more about the hard X-ray emission of 1E~1048.1$-$5397, but this will require several megaseconds
of exposure.  Of the persistently X-ray bright magnetars, 1E~1048.1$-$5937 is the faintest in the soft band;
our faint 10--20 keV detection, in spite of the lengthy {\it NuSTAR} exposure, demonstrates that {\it NuSTAR}
hard-band detections of fainter magnetars -- unless they are in outburst -- will be challenging.

\acknowledgments

We thank Daniel Stern for helpful comments.
This work was supported under NASA Contract No. NNG08FD60C, and made use of data from the {\it NuSTAR} 
mission, a project led by the California Institute of Technology, managed by the Jet Propulsion Laboratory, 
and funded by the National Aeronautics and Space Administration. We thank the {\it NuSTAR} Operations, 
Software and Calibration teams for support with the execution and analysis of these observations. This
research has made use of the {\it NuSTAR} Data Analysis Software (NuSTARDAS) jointly developed by the
ASI Science Data Center (ASDC, Italy) and the California Institute of Technology (USA).  
Part of this work was performed under the auspices of the U.S. Department of Energy by Lawrence Livermore National Laboratory under Contract DE-AC52-07NA27344.
RFA receives support from an NSERC  Alexander Graham Bell Canada Graduate Scholarship.
JAK was supported by supported by NASA contract NAS5-00136.
VMK receives support from an NSERC Discovery Grant and Accelerator Supplement, from the Centre de Recherche en Astrophysique du Qu\'ebec, an R. Howard Webster Foundation Fellowship from the Canadian Institute for Advanced Study, the Canada Research Chairs Program and the Lorne Trottier Chair
in Astrophysics and Cosmology.  AMB was supported by NASA grants NNX-10-AI72G and NNX-13-AI34G.

\end{document}